\documentclass[graybox]{svmult}

\usepackage{natbib}

\usepackage{mathptmx}       
\usepackage{helvet}         
\usepackage{courier}        
\usepackage{type1cm}        
\usepackage{makeidx}         
\usepackage{graphicx}        
\usepackage{multicol}        
\usepackage[bottom]{footmisc}

\makeindex             

\newcommand{\HII}{H$\,${\sc ii}}
\newcommand{\HI}{H$\,${\sc i}}
\newcommand{\rd}{$R_{\rm D}$}
\newcommand{\rbr}{$R_{\rm Br}$}
\newcommand{\rtf}{$R_{25}$}
\newcommand{\coldens}{atoms\,cm$^{-2}$}

\begin{document}

\title*{Outskirts of Nearby Disk Galaxies: Star Formation and Stellar
Populations}
\author{Bruce G. Elmegreen and Deidre A. Hunter}
\institute{Bruce G. Elmegreen \at IBM Research Division, T.J. Watson Research
Center, 1101 Kitchawan Road, Yorktown Heights, NY 10598 USA,
\email{bge@us.ibm.com}, \and Deidre A. Hunter \at Lowell Observatory, 1400 West
Mars Hill Road, Flagstaff, Arizona 86001, USA, \email{dah@lowell.edu}, }
%
%
\maketitle


\abstract{The properties and star formation processes in the far-outer disks of
nearby spiral and dwarf irregular galaxies are reviewed. The origin and structure
of the generally exponential profiles in stellar disks is considered to result from
cosmological infall combined with a non-linear star formation law and a history of
stellar migration and scattering from spirals, bars, and random collisions with
interstellar clouds. In both spirals and dwarfs, the far-outer disks tend to be
older, redder and thicker than the inner disks, with the overall radial profiles
suggesting inside-out star formation plus stellar scattering in spirals, and
outside-in star formation with a possible contribution from scattering in dwarfs.
Dwarf irregulars and the far-outer parts of spirals both tend to be gas dominated,
and the gas radial profile is often non-exponential although still decreasing with
radius. The ratio of H$\alpha$ to far-UV flux tends to decrease with lower surface
brightness in these regions, suggesting either a change in the initial stellar mass
function or the sampling of that function, or a possible loss of H$\alpha$
photons.}

\section{Introduction}
\label{intro}

The outer parts of galaxies represent a new frontier in observational astronomy at
the limits of faint surface brightness\index{surface brightness}. We know little about these regions except
that galaxies viewed deeply enough can usually be traced out to 10 stellar scale
lengths or more, without any evident edge. We do not know in detail
how the stars and gas got there and whether stars actually formed there or just
scattered from the inner parts. Neither do we know as much as we'd
like about the properties, elemental abundances, scale heights and kinematics of
outer disk stars except for a limited view in the Milky Way (e.g.,
\citealt{Bovy16}) and the Andromeda galaxy (\citealt{Dalcanton12}). Yet the outer parts
of disks are expected to be where galaxy growth is occurring today,
and where the left-over and recycled cosmological gas accretes or gets stored for
later conversion into stars in the inner disk (\citealt{Lemonias11},
\citealt{Moffett12}). The outer parts should also show the history of a galaxy's
interactions with other galaxies, as the orbital time is relatively long. A high
fraction of outer disks are lopsided\index{lopsided} too, correlating with the stellar mass
fraction in the outer parts (i.e., the ratio of the stellar mass to the total from
the rotation curve; \citealt{Zaritsky13}), perhaps because of uneven accretion,
interactions, or halo sloshing (\citealt{Ghosh16}). This chapter reviews disk
structure, star formation and stellar populations in the outer parts of nearby
galaxies. General properties of these outer disks are in Sect.~\ref{collapse} to
\ref{sfsection}, and a focus on dwarf irregular galaxies (dIrrs) is in Sect.~\ref{sec:deidre_dwarfs}.  The observational difficulties in observing
the faint outer parts of disks are discussed in other chapters in this volume.

\section{Outer disk Structure from Collapse Models of Galaxy Formation}
\label{collapse}

A fundamental property of galaxy disks is their exponential or piece-wise
exponential radial light profile\index{exponential light profile} (\citealt{dev59}). \citet{Freeman70} noted that
this profile gives a distribution of cumulative angular momentum versus radius that
matches that of a flattened uniformly rotating sphere (\citealt{Mestel63}), but
this match is only good for about four disk scale lengths. The problem is that an
exponential disk has very little mass and a lot of angular momentum\index{halo angular momentum} in the
far-outer parts, unlike a power-law halo which has both mass and angular momentum
increasing with radius in proportion (\citealt{Efstathiou00}). Nevertheless,
observations show some disks with 8 to 10 scale lengths (\citealt{Weiner01},
\citealt{BlandHawthorn05}, \citealt{Grossi11}, \citealt{Hunter11},
\citealt{RadburnSmith12}, \citealt{Vlajic11}, \citealt{Barker12} ,
\citealt{Hunter13}, \citealt{mihos13}, \citealt{vandokkum14}). These large extents
compared to the predicted four scale lengths from pure collapse models
need to be explained (\citealt{Ferguson01}).

Thus we have a problem: if the halo collapses to about four scale lengths\index{scale length} in a disk,
then how can we get the observed eight or more scale lengths in the stars that
eventually form? The answer may lie with the conversion of incoming gas into stars.
In a purely gaseous medium, interstellar collapse proceeds at a rate
per unit area that is proportional to the square of the mass column density,
$\Sigma_{\rm gas}$ (\citealt{Elmegreen15}). One factor of $\Sigma_{\rm gas}$
accounts for the amount of fuel available for star formation and the other factor
accounts for the rate of conversion of this fuel into stars. This squared
Kennicutt-Schmidt law converts four scale lengths of primordial gas into eight scale
lengths of stars after they form (Sect. \ref{sfsection}). Stellar scattering from
clouds and other irregularities could extend or smooth out this exponential further
(\citealt{Elmegreen13}, \citealt{Elmegreen16b}).

There is an additional observation in \citet{Wang14} that in local gas-rich galaxies, the
outer gas radial profiles are all about the same when scaled to the radius where
$\Sigma_{\rm HI} = 1\,M_\odot$ pc$^{-2}$. \citet{Bigiel12} found a similar universality
to the gas profile when normalized to $R_{25}$, the radius at 25 magnitudes per square
arcsec in the $V$ band. \citet{Wang14} found that the ratio of the radius at 1 $M_\odot$
pc$^{-2}$ to the gaseous scale length in the outer disk\index{outer disk gas} is about four, the same as the
maximum number of scale lengths in a pure halo collapse. This similarity may not be a
coincidence (Sect. \ref{sfsection}).

Cosmological simulations now have a high enough resolution to form individual galaxies
with reasonable properties (\citealt{Vogelsberger14}, \citealt{Schaye15}). Zoom-in models\index{zoom-in models}
in a cosmological environment show stellar exponential radial profiles in these
galaxies (\citealt{Robertson04}) even though specific angular momentum is not preserved
during the collapse and feedback moves substantial amounts of gas around, especially for
low-mass galaxies (\citealt{ElBadry16}). For example, \citet{AumerWhite13} ran models
with rotating halo gas aligned in various ways with respect to the dark matter symmetry
axis. They found broken exponential disks with a break radius\index{break radius} related to the maximum
angular momentum of the gas in the halo, increasing with time as the outer disk cooled
and formed stars. Star formation is from the inside-out.  Angular momentum was
redistributed through halo torques, but still the disks were approximately exponential.
\citet{Aumer13} further studied 16 simulated galaxies with various masses. All of them
produced near-exponential disks.

In a systematic study of angular momentum, \citet{Herpich15} found a transition from
exponentials with up-bending outer profiles\index{up-bending outer profiles} (Type III---Sect.~\ref{types}) at low
specific angular momentum\index{specific angular momentum} ($\lambda$) to Type I (single exponential) and Type II
(down-bending\index{down-bending outer profiles} outer parts) at higher $\lambda$. An intermediate value of $\lambda=0.035$,
similar to what has been expected theoretically (\citealt{Mo98}), corresponded to the
pure exponential Type I. The reason for this change of structure with $\lambda$ was that
collapse at low spin parameter produces a high disk density in a small initial radius,
and this leads to significant stellar scattering and a large redistribution of mass to
the outer disk, making the up-bending Type III. Conversely, large $\lambda$ produces a
large and low-density initial disk, which does not scatter much and nearly preserves the
initial down-bending profile of Type II.

\section{Outer Disk Structure: Three Exponential Types}
\label{types}

\begin{figure}[t!]
\includegraphics[width=\textwidth]{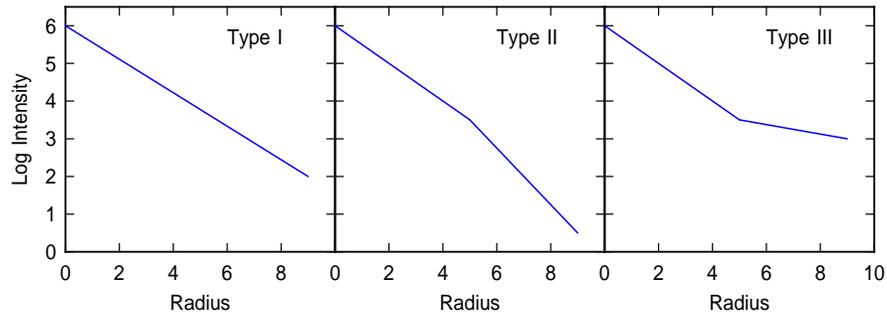}
\caption{Three types of exponential or piece-wise exponential profiles
}
\label{knapen_review_3types}
\end{figure}

Galaxy radial profiles are often classified as exponential Types I, II or III\index{exponential Types I, II, III}
according to whether the outer parts continue with the same scale length as the
inner parts, continue with a shorter scale length (i.e., bend down a little) or
continue with a larger scale length (bend up a little), respectively (Fig.
\ref{knapen_review_3types}). The Sersic profile\index{Sersic profile} with $n=1$ corresponds to Type I;
the other types do not have a constant Sersic index. For a review, see
\citet{vanderKruit01}, and for early surveys, see \citet{Pohlen06} and
\citet{Erwin08}.

\citet{Gutierrez11} determined the proportion of the three exponential profile
types for barred and non-barred galaxies of various Hubble types,
including 183 large local face-on galaxies from three separate studies. For S0 and
earlier, the three profile types are nearly evenly divided. For Sab to Sbc, Types
II and III are about equal and Type I becomes relatively rare (10\%). For Scd to
Sdm, Type II dominates with $\sim80$\% of the total. \citet{Herrmann13} continued
this study to dIrrs and blue compact dwarfs (BCDs, see also
Sect. \ref{subsec:deidre_breaks}). dIrr galaxies are dominated (80\%) by Type IIs,
while BCDs have steep inner parts from a starburst and are usually Type III.

A general caution should be mentioned about possible contamination at faint light levels
from scattered light\index{scattered light}. \citet{Sandin15} showed $R$ and $I$ band radial profiles for
NGC 4102 that were fit to a single exponential model with a broad point spread function
from the instrument. The usual Type III profile for this galaxy turned into a Type I when
the outer excess was corrected for the instrumental profile.

\section{Outer Disk Stellar Populations: Colour and Age Gradients}

Radial profiles become more complex when changes in stellar colours and
ages\index{galaxy colour gradients}\index{galaxy age gradients} are
considered. \citet{Bakos08} noted that Type II light profiles tend to correspond to
U-shaped $B-V$ colour profiles, which means that the inner part of the disk gets
bluer with radius at first, and then the outer part of the disk gets red again.
This colour change presumably corresponds to a change in the mass-to-light ratio,
with large ratios in the outer parts. Then the down-bending Type II in a light
profile tends to straighten out and become Type I in a mass profile.
That is, the outer red trend gives an increasing mass-to-light ratio, causing an
increasing conversion factor from surface brightness to mass surface density. The
red outer parts could be from old stars that scattered\index{scattered stars} there from the inner regions
(\citealt{Roskar08}), as distinct from the common model of inside-out growth for
spiral galaxies.  A larger survey recently confirmed this result. \citet{Zheng15}
included 700 galaxies using deep images from the Pan-STARRS survey. The average
$g$-band (peak at $5150$\,\AA) light profile was the down-bending Type II for low-mass
galaxies ($<10^{10}\,M_\odot$) and slightly less bent for high mass galaxies
($<10^{10.5}\,M_\odot$), as usual, and the average $g-i$ colour profiles ($i$ band
peaks at $7490\AA$) were U-shaped to various degrees, so the average mass profile
became Type I for all galaxy masses. \citet{MunozMateos15} made average radial
profiles separated into eight mass bins for $\sim2400$ galaxies using 3.6\,$\mu$m emission
from the {\it Spitzer} Survey of Stellar Structure in Galaxies. Such long wavelength
emission is a nearly direct probe of galaxy mass, although there is some PAH
emission from dust in it too. All masses showed Type II profiles on average, with a
straighter trend like Type I from a central bulge in the more massive galaxies.
This suggests that the mass profile for most galaxies is not exactly Type I, but
still tapers off more steeply in the outer parts, beyond 1\,kpc for low mass
galaxies ($<10^9\,M_\odot$) and beyond 10\,kpc for high mass galaxies
($>10^{10.5}\,M_\odot$).

The most telling observations are of stellar age gradients\index{galaxy colour
gradients}\index{galaxy age gradients} because colour gradients
can be from a mixture of age gradients and metallicity gradients.
\citet{Roediger12} determined radial age profiles from photometry and stellar
population models of 64 Virgo cluster disk galaxies. They found U-shaped age
profiles\index{U-shaped profiles} in 15\% of Type I's, and also in 36\% of both Types II and III.  In
one-third of all exponential types, the age increased steadily with radius.
\citet{Yoachim12} found about the same mixture of age profiles, measuring ages from
spectra in 12 galaxies. \citet{Dale16} determined star formation
histories for 15 nearby galaxies with masses in the range $10^8\,M_\odot$ to
$10^{11}\,M_\odot$ using ultraviolet and infrared data; they also found U-shaped age
profiles. These results imply that outer disks generally have old stars, although
most also still have some star formation.

A recent integral field unit survey of 44 nearby spiral galaxies
(CALIFA) by \citet{RuizLara16} also found U-shaped age profiles in Types I and II
when the stars were weighted by brightness, as would be the case from integrated
spectra or photometry. This is in agreement with the previous surveys mentioned
above. The galaxies were observed beyond their break radii or for at
least three scale lengths. In contrast, \cite{RuizLara16} found constant age profiles
when the stars were weighted by mass. They suggested that the entire disk formed
early with star formation stopping in the inner parts first, and then quenching
from inside-out. This is unlike cosmological simulations that have the outer disk
form more slowly than the inner disk, and also unlike models where the outer disk
stars migrate there from the inner disk.

\citet{Watkins16} viewed three nearby spirals with very deep images,
covering a range of about 10 magnitudes in surface brightness for $B$ band. They
found smooth and red stellar distributions with no spiral arms in the
far-outer disks. For the typical colour of $B-V \sim 0.8$ mag in the outer
parts, and from a lack of FUV light, they concluded
that the star formation rate (SFR) had to be less than
$3-5\times10^{-5}\,M_\odot$\,pc$^{-2}$\,Myr$^{-1}$. This seemed to be too low for
continuous star formation and disk building, suggesting some radial migration.
However, the lack of spiral arms makes the usually invoked churning mechanism
(\citealt{Sellwood02,Roskar08,Berrier15}) inoperable. Churning is a
process of stellar migration back and forth around corotation. Perhaps stellar
scattering\index{stellar
scattering}\index{stellar
churning} off local gas irregularities makes the outer
exponential structure (\citealt{Elmegreen13}).

\section{Mono-age Structure of Stellar Populations}

Age profiles\index{mono-age structure} in galaxy disks can be viewed in another way too. A series of galaxy
simulations have looked at the distributions of stars of various ages in the final model.
For example, \citet{Bird13} did a simulation of the Milky Way and found that older stars
in the present-day disk have shorter radial scale lengths and thicker perpendicular
scale heights than younger stars. Other mono-age population studies of simulated disks
are in \citet{SanchezBlazquez09}, \citet{Stinson13}, \citet{Martig14}, \citet{Minchev15}
and \citet{Athanassoula16}, giving the same result.

\citet{Bovy16} found structure related to this in the Milky Way using 14700 red clump
stars. Higher metallicity populations are more centrally concentrated than lower
metallicity populations (not considering the $\alpha$-enhanced ``thick disk'' component).
Each narrow metallicity range tends to have a maximum surface density of stars at a
particular radius where the disk has that average metallicity.  Plus, each
mono-metallicity population has a perpendicular scale height that increases with radius,
producing a flare\index{flare}.

The correspondence between metallicity and peak surface density for a population of stars
suggests that star formation, feedback, halo recycling, and other processes establish an
equilibrium metallicity\index{equilibrium metallicity} in a region that depends primarily on local conditions, such as
the local mass surface density (\citealt{Bovy16}). Stellar migration\index{stellar migration} then broadens this
distribution to produce the observed total profiles. This local equilibrium concept is
consistent with the results of \citet{RosalesOrtega12}, who found for 2000 H{\sc ii} regions in
nearby galaxies that metallicity depends mostly on stellar
mass surface density, as determined from photometry. \citet{Bresolin15} present a similar
result: that the metallicity gradients\index{metallicity gradients} in galaxies are all the same when expressed in
units of the disk scale length.

\section{Outer Disk Structure: Environmental Effects and the Role of Bulges and Bars}

Environment may also affect outer disk structure. \citet{Younger07} showed that prograde
minor mergers can drive mass inward and outward, creating a Type III profile.
{\citet{Borlaff14} also suggested that Type III S0 galaxies can result from a merger\index{mergers}.
This is consistent with observations in \citet{Erwin12} that S0 galaxies in Virgo have
proportionally more Types I and III, suggesting that interactions or mergers have been
important. Erwin et al. also found that bars\index{bars} have little effect on the proportion of
exponential types. \citet{Athanassoula16} simulated a gas-rich major merger and showed
that it formed an exponential disk in the final system.

On the other hand, \citet{Maltby12} measured the $V$-band radial profiles of 330
galaxies observed with {\it Hubble Space Telescope}  over a
half-degree field surrounding a galaxy supercluster at redshift 0.165. They found
no dependence on environment\index{environment}, cluster versus field, for the ratio of the outer to
the inner disk scale length or the outer scale length itself. \citet{Head15} got a
similar result looking at S0 galaxies in the Coma cluster; using a
profile decomposition algorithm to remove the bulge, they found that bars are
important for disk structure, correlating with Types II and III (contrast this with
the \citealt{Erwin12} result above), but that location in the cluster is not
important. According to \citet{Head15}, the relative proportion of Types I, II, and
III is the same in the core, at intermediate radii, and in the outskirts of Coma
(in fact, most of the S0 galaxies were Type I).

Some of the appearance of Type III could be from a bright halo\index{halo} or extended bulge and not
from stars in the disk (\citealt{Erwin05}). \citet{Maltby15} suggested that half of the
S0 Type III structures in various environments come from extended bulge light, although
this fraction is only 15\% in later Hubble type spirals. This implies that disk fading
can make an S0 from a spiral, preserving the scale length. Simulations by
\citet{Cooper13} also found that the outer stellar structure can be in a halo and
not a disk, as a result of mergers.

Bars and spirals seem to be important in determining the break radius\index{break radius} for down-bending
(Type II) exponentials. \citet{MunozMateos13} suggested that the break radius for Type
II's is either at the outer Lindblad resonance (OLR) of a bar or the OLR of a spiral that
is outside of a bar. The spiral and bar are assumed to have their pattern speeds\index{pattern speed} in a
resonance\index{resonance} with the inner 4:1 resonance of the spiral at the corotation radius of the bar
(see \citealt{Pohlen06} and \citealt{Erwin08}). \citet{Laine14} also found that bars and
spirals are important: 94\% of Type II breaks are associated with some type of feature;
48\% are in early type galaxies with an outer ring or pseudoring; 8\% are with a lens,
assumed to be the OLR of a bar, and if there is no outer ring, then the breaks are at 2
times the radius of an inner ring (this being the ratio of radii for outer and inner ring
resonances); 14\% are in late type galaxies associated with an end to strong star
formation, and 24\% are at the radius where the spiral arms end. For Type III breaks
studied by \citet{Laine14}, 30\% are associated with inner or outer lenses or outer
rings.

\section{Outer Disk Structure: Star Formation Models}
\label{sfsection}

The\index{star formation} outer disks of spiral galaxies and most dIrrs are
dominated by gas in an atomic form, and not stars. Because stars form in molecular
gas, it is difficult to observe directly how stars form in these regions. Moreover,
outer disks and dIrrs tend to be stable by the Toomre $Q$ condition
(\citealt{EH15}). Nevertheless, star formation usually looks normal
there, forming clusters and associations at low density
(\citealt{Melena09,Hunter16}), although it may stop short of the full extent of
the gas disk (see, for example, Fig.~\ref{fig-spirals}.)

\begin{figure}[t!]
\centering
\includegraphics[scale=0.9]{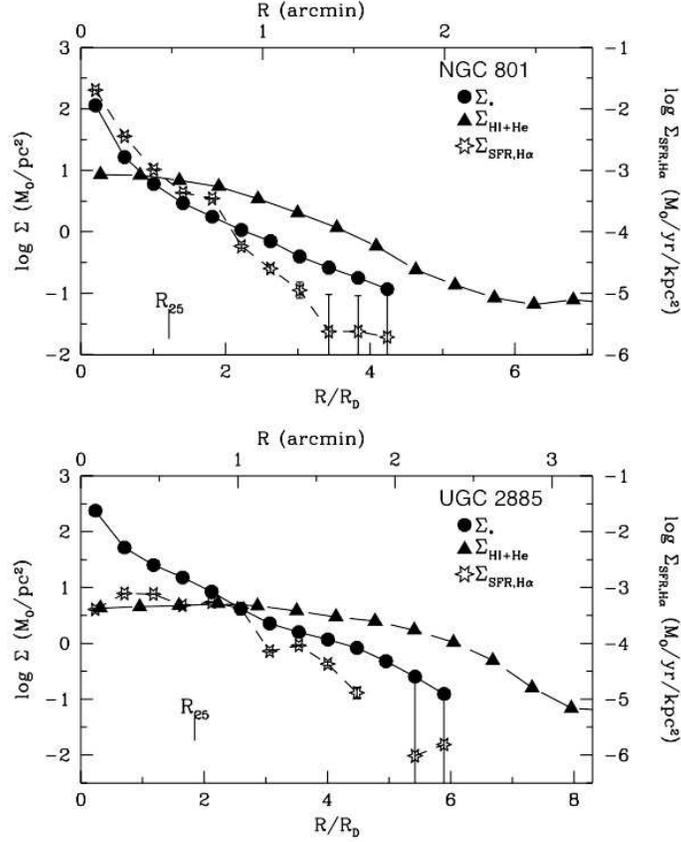}
\caption{Stellar mass surface density $\Sigma_*$, \HI$+$He surface density $\Sigma_{\rm
HI+He}$, and SFR density $\Sigma_{\rm SFR, H\alpha}$ plotted as a function of
radius for two very luminous ($M_V=-22$ to $-23$) Sc-type spiral galaxies, NGC~801\index{NGC 801}
and UGC~2885\index{UGC 2885}. The radius is normalized to the optical $V$-band disk scale length
$R_{\rm D}$. The gas and stellar mass surface densities have been corrected to face-on.
The logarithmic interval is the same for all three quantities, but the SFR zero
point is different. Adapted from \citet{Hunter13} } \label{fig-spirals}
\end{figure}

\citet{Krumholz13} formulated a model for star formation in these conditions that
considers the existence of a two-phase atomic medium\index{two-phase medium} (i.e., a warm neutral medium in
pressure equilibrium with a cool neutral medium) and the molecular fraction\index{molecular fraction} in such a
medium. He then assumed that star formation occurs in the molecular medium at a rate
given by an efficiency per unit free fall time\index{efficiency per unit free fall time}, $\epsilon_{\rm ff}\sim0.01$, times the
molecular mass divided by the free fall time in the molecular gas:
\begin{equation}
\Sigma_{\rm SFR}=\epsilon_{\rm ff}\Sigma_{\rm mol}/t_{\rm ff,mol}.
\end{equation}
The free fall time depends on the molecular cloud density, which for outer disks in their
model, depends on the molecular cloud mass and a fiducial value of the molecular cloud
surface density, $\Sigma_{\rm GMC}=85\,M_\odot$\,pc$^{-2}$. The molecular cloud mass was
taken to be the turbulent Jeans mass in the interstellar medium (ISM), $M_{\rm
GMC}=\sigma^4/(G\Sigma_{\rm gas})$ for turbulent speed $\sigma$ and average ISM gas
surface density $\Sigma_{\rm gas}$.  For inner disks, the free fall time was taken to be
the value for an average disk density where the Toomre $Q$ parameter equals unity. To span the
inner and outer regions, the minimum of these two free fall times was used.

One uncertainty in the \citet{Krumholz13} model is the assumption that a two-phase
medium is present, because it need not be present everywhere in the outer disk. But
this assumption seems reasonable for star formation because cool gas greatly
facilitates cloud formation (\citealt{Elmegreen94}, \citealt{Schaye04},
\citealt{Forbes16}).  A second assumption is that the SFR is given
only by the molecular gas mass and density, and that this is related to the total
density by the molecular fraction, which depends on the ratio of the radiation
field to the cool cloud density. In this model, molecule formation is calculated
separately as a precursor to star formation, and then whatever is calculated for
molecules is used to determine the SFR.

Another model considers that the average SFR is determined by the
average ISM dynamics and that molecule formation is incidental, i.e., molecule
formation happens along the way but it is not a limiting factor. Star formation in
predominantly atomic gas has been predicted by \citet{Glover12} and
\citet{Krumholz12} and suggested by observations in \citet{Michalowski15} and
\citet{Elmegreen16}. In this model, the gas mass available for star formation is
the total gas mass in all forms, even atomic gas, and the free fall rate of this
gas is given by the average midplane density, regardless of molecular content
(\citealt{Elmegreen15}). Cool clouds are still required so the ISM cannot be purely
warm phase. Also, because molecular hydrogen is slow to form at low density
(\citealt{MacLow12}), there could be a substantial fraction of H$_2$ in stagnant,
diffuse clouds without significant CO emission and with little connection to star
formation (\citealt{EH15}). Such a diffuse H$_2$ medium\index{diffuse molecular medium} was found in simulations by
\citet{Hu16} and \citet{Safranek16} but has not been observed yet. These diffuse
H$_2$ clouds, along with more atom-rich clouds, would presumably come together
during localized ISM collapse as a precursor to star formation. In this model the
SFR per unit area is given by
\begin{equation}
\Sigma_{\rm SFR} = \epsilon_{\rm ff} \Sigma_{\rm gas} / t_{\rm ff,gas}\label{sfl0}\end{equation}
for midplane free-fall time $t_{\rm ff}$ at the density $\rho = \Sigma_{\rm gas}/(2H)$
and scale height $H = \sigma^2 / ( \pi G \Sigma_{\rm gas})$.
The result is
(\citealt{Elmegreen15})
\begin{equation}
\Sigma_{\rm SFR} = \epsilon_{\rm ff} ( 4/3^{1/2} ) ( G / \sigma ) \Sigma_{\rm gas}^2
= 1.7 \times 10^{-5} (\Sigma_{\rm gas}/[1\,M\odot/{\rm pc}^2])^2 (\sigma/6\,{\rm km\, s}^{-1})^{-1}
\label{sfl}
\end{equation}
where $\sigma$ is the gas velocity dispersion and $\epsilon_{\rm ff}\sim1$\% is the
efficiency of star formation per unit free fall time.

Both of the above models compare well with observations (\citealt{Krumholz13},
\citealt{Elmegreen15}).

\citet{Hu16} simulated dwarf galaxies with a chemical model to form H$_2$, CO and
other molecules, cloud self-shielding from radiation, and a SFR given
by Eq.~(\ref{sfl0}) at a threshold density\index{threshold density} of 100\,cm$^{-3}$ and a temperature
less than 100\,K. They found that $\Sigma_{\rm SFR}$ decreases faster than
$\Sigma_{\rm gas}$ but not because of a flare (the extra $\Sigma_{\rm gas}$ factor
in Eq.~\ref{sfl}). Rather, $\Sigma_{\rm SFR}$ follows the cold gas with a rate
that scales directly with the cold gas fraction, i.e., a linear law, and this cold
gas fraction decreases with radius. The linear law in their model is because of the
assumed constant threshold density. Most star-forming gas is close to this fixed
density, so the characteristic dynamical time is the fixed value at this density.

This point about a fixed density is similar to the explanation for the linear star
formation law in \citet{Elmegreen15}, where it was pointed out that if CO, HCN, and
other star formation tracers emit mostly at their fixed excitation density\index{excitation density}, as
determined by the Einstein $A$ coefficient, then the effective free fall time is
the fixed value at this density. The SFR then scales only with the
amount of gas at or above this observationally selected density.  The fixed density
in this case is not because of an assumption about a star formation threshold, as
there is no threshold in the \citet{Elmegreen15} model. There is just a continuous
collapse of ISM gas at a rate given by the midplane density, and a feedback return
of the dense gas to a low density form.

The existence of a fixed threshold density for star formation is something to be tested
observationally. \citet{Elmegreen15} suggest there is no threshold density because clouds
are strongly self gravitating when $\pi G\Sigma_{\rm cloud}^2>P_{\rm ISM}$ for cloud
surface density $\Sigma_{\rm cloud}$ and ambient pressure $P$. Because the interstellar
pressure varies with the square of the total surface density of gas and stars inside the
gas layer, there is a large range in pressure over several exponential scale lengths in a
galaxy disk---a range that may exceed a factor of 100 for dIrrs, and 1000 for
spirals. Thus if there is a threshold for star formation, the above equation suggests
that it might be
\begin{equation}
\Sigma_{\rm cloud}>\left(P/\pi G \right)^{1/2},
\end{equation}
in which case it should vary with radius.

We return to a point made in Sect.~\ref{intro} about the gas surface density profile in
the far-outer regions of gas-rich galaxies. \citet{Wang14} noted that the gas exponential
scale length beyond the radius $R_1$, where $\Sigma_{\rm gas}=1\,M_\odot$ pc$^{-2}$, is
always about 0.25 times this radius. We can see this here also from Eq.~(\ref{sfl}),
which states that
\begin{equation}
\Sigma_{\rm SFR} = 2\times10^{-5}\,M_\odot\,{\rm pc}^{-2}\,{\rm Myr}^{-1}\,{\rm at}\,
\Sigma_{\rm gas}=1\,M_\odot\,{\rm pc}^{-2}.
\end{equation}
After a Hubble Time of $10^4$\,Myr, $\Sigma_{\rm stars}$ is approximately $0.2\,M_\odot$\,pc$^{-2}$.
According to the average disk mass profiles in \citet{Zheng15}, this outer
stellar surface density is lower than that at the disk centre by $10^{-3.5}$ on average,
which represents eight scale lengths in stars. But eight scale lengths in stars is four scale
lengths in gas for Eq.~(\ref{sfl}). Thus the radius at 1\,$M_\odot$\,pc$^{-1}$ is
about four times the scale length in the gas, as observed further out by \citet{Wang14}.

\section{The Disks of Dwarf Irregular Galaxies}
\label{sec:deidre_dwarfs}

Dwarf irregular galaxies\index{dwarf irregular galaxies} are like the outer parts of spiral galaxies in terms of
gas surface density, SFR, and gas consumption time. Tiny dIrrs
have extended exponential disks as well. For example,
\citet{Saha10} traced the Large Magellanic Cloud\index{Large Magellanic Cloud} (LMC) to 12\,\rd, an
effective surface brightness of 34\,mag\,arcsec$^{-2}$ in $I$, and \citet{Sanna10} found stars in IC~10\index{IC~10} to $\sim10$\,\rd.
\citet{Bellazzini14} detected stars associated with Sextans A and Sextans B\index{Sextans A}\index{Sextans B} to
6\,\rd,
and Hunter et al.\ (2011) measured surface brightness profiles in four nearby dIrrs and
one BCD to 29.5\,mag\,arcsec$^{-2}$ in $V$, corresponding to $3-8$\,\rd.
These extended stellar disks represent extreme galactic environments for star formation
and are potentially sensitive probes of galaxy evolutionary processes, and yet they are
relatively unexplored. In this Section we examine what is known about outer disks of dIrr
galaxies.

\subsection{Radial Trends}
\label{subsec:deidre_trends}

\subsubsection{The Gas Disk}
\label{subsubsec:deidre_hi}

The \HI\ gas often dominates the stellar component of dIrr galaxies, both in extent and
mass. How much further the gas extends compared to the stars was demonstrated by
\citet{Krumm84} for DDO~154\index{DDO~154} where the \HI\ was traced to 8\,\rtf\ at a column density of
$2\times10^{19}$\,\coldens\ ($0.22\,M_\odot$ pc$^{-2}$). In the LITTLE THINGS sample of 41
nearby ($<10.3$ Mpc), relatively isolated dIrrs (\citealt{Hunter12}), most systems have
gas extending to $2-4$\,\rtf\ or $3-7$\,\rd\ at that same (face-on) column density. Some spiral
galaxies also have extended \HI; \citet{Portas10} found that the Sbc galaxy NGC~765\index{NGC~765}, for
example, has gas extending to 4\,\rtf. Large holes (up to 2.3\,kpc diameter) are also
sometimes found in the gas beyond 2\,\rd\ (\citealt{Dopita85}, R.N. Pokhrel, in
preparation).

In most dIrrs, the galaxy is gas-dominated\index{gas-dominated} and becomes increasingly gas-rich with radius
(Fig.~\ref{fig-startogas}). This implies a decreasing large-scale star formation
efficiency (\citealt{Leroy08}, \citealt{Bigiel10}). The lack of sharp transitions in the
star-to-gas ratio, including at breaks in the optical exponential surface brightness
profiles, suggests that the factors dominating the drop in star formation with radius are
changing relatively steadily.

\begin{figure}[t!]
\centering
\includegraphics[width=\textwidth]{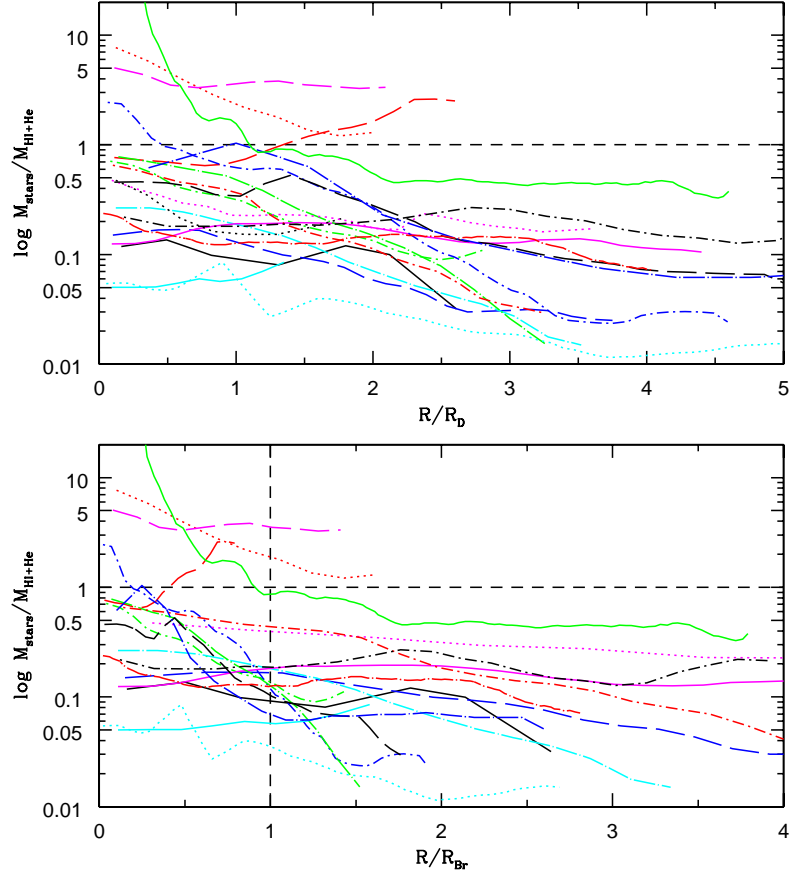}
\caption{Azimuthally averaged stellar mass to gas mass ratios as a function of radius
normalized to the disk scale length ({\it top}) and radius at which the $V$-band surface
brightness profile changes slope \rbr\ ({\it bottom}). These galaxies are from the LITTLE
THINGS sample with stellar mass profiles determined by \citet{Zhang12} and gas mass
profiles from Hunter et al.\ (2012) } \label{fig-startogas}
\end{figure}

The gas surface density drops off with radius usually in a non-exponential fashion.  In
Sect.~$2-7$, we have approximated the radial gas profiles of spiral galaxies as
exponentials. However, especially in dIrrs, the gas profiles are rarely pure
exponentials, and since dIrrs are gas-dominated, the shape of the gas profile is crucial.
Note, however, that in dIrrs, the radial {\it stellar} profiles are usually exponential
in shape. In a sub-sample of the THINGS spirals (\citealt{Walter08}), \citet{Portas10}
found that the gas is approximately constant at $5-10\times10^{20}$\,\coldens\ and then
drops off rapidly. A Sersic\index{Sersic index for gas} function fits the profiles with indices of $n=0.14-0.22$. For
comparison, an exponential disk has an $n$ of 1.0. In five THINGS dIrrs the gas density
dropped more shallowly with radius, and the distribution of $n$ peaked around 0.3. For
the LITTLE THINGS sample of dwarfs, the shape of the \HI\ radial profiles varied from
galaxy to galaxy, and $n$ varied from 0.2 to 1.65 with most having values $0.2-0.8$. The
lack of correlations between the \HI\ profile index $n$ and characteristics of the
stellar disk suggest that the role of the gas distribution in determining the stellar
disk properties is complex.

\subsubsection{The Stellar Disk}
\label{subsubsec:deidre_stars}

\citet{Zhang12} performed spectral energy distribution\index{spectral energy distribution} fitting to
azimuthally-averaged surface photometry of the LITTLE THINGS galaxies. The fitting
included up to 11 passbands from the FUV to the NIR. From these fits they
constructed SFRs as a function of radius over three broad timescales:
100\,Myr, 1\,Gyr, and galaxy lifetime. Zhang et al found that the bulk star formation
activity has been shrinking with radius over the lifetime of dwarf galaxies, and
they adopted the term ``outside-in'' disk growth\index{outside-in disk growth in dwarf irregulars}. Although Zhang et al found that
``outside-in'' disk growth applied primarily to dIrrs with baryonic masses $<10^8\,
M_\odot$, \citet{Gallart08} and \citet{Meschin14} found the same phenomenon in the
LMC, a more massive irregular galaxy. Similarly, \citet{Pan15} suggested
from colour profiles that the same process is occurring in a large
sample of Sloan Digital Sky Survey galaxies with stellar masses up to $10^{10}\,M_\odot$. This
outside-in disk growth is in contrast to the inside-out disk growth\index{inside-out disk growth in spirals} identified in
spirals (\citealt{White91}, \citealt{Mo98}, \citealt{MunozMateos07},
\citealt{Williams09}, but see \citealt{RuizLara16}).


\begin{figure}[t!]
\centering
\includegraphics[scale=0.8]{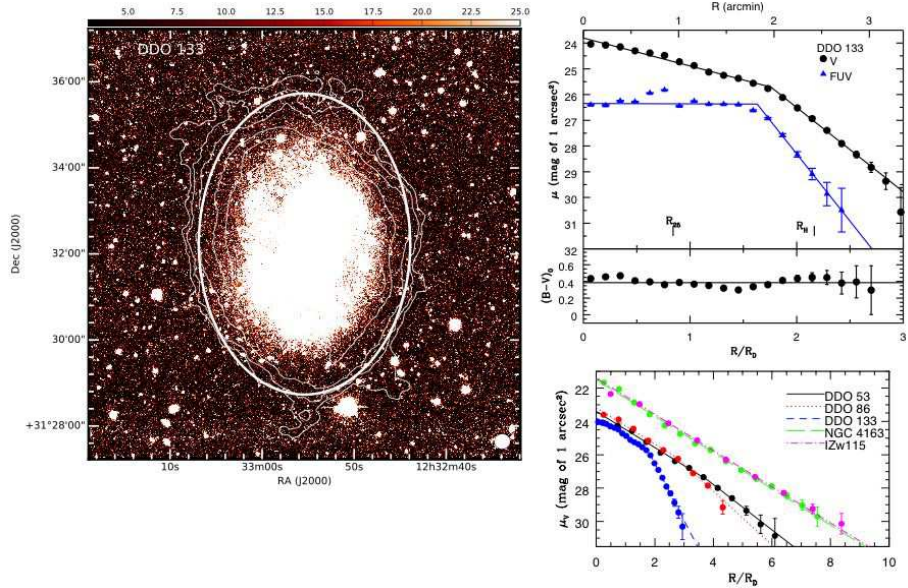}
\caption{{\it Left}: $V$-band image of DDO 133 from \citet{Hunter11}. The white ellipse marks the
extent of the galaxy measured to 29.5 mag arcsec$^{-2}$. The white contours trace
column densities of 5, 30, 100, 300, 500, 1000, and $3000\times10^{18}$\,atoms\,cm$^{-2}$
in \HI\ (Hunter et al.\ 2012). {\it Top right}: Surface photometry in $V$, FUV,
and $B-V$ as a function of radius normalized to the disk scale length $R_{\rm D}$. {\it Bottom
right}: $V$-band surface photometry of all five galaxies from \citet{Hunter11}. IZw
115 is a BCD; the rest are dIrrs } \label{fig-deidre_d133}
\end{figure}

\citet{Hunter11} carried out an ultra-deep imaging program on four nearby dIrrs and
one BCD. They measured surface photometry in this sample to 29.5\,mag arcsec$^{-2}$
in $V$, and also obtained deep $B$ images of three of the galaxies and deep FUV and
NUV images with the {\it Galaxy Evolution Explorer} ({\it GALEX},
\citealt{Martin05}). Fig.~\ref{fig-deidre_d133} shows the $V$-band image and
photometry of DDO 133, illustrating what they found. What does a surface brightness
of 29.5\,mag\,arcsec$^{-2}$ mean? In DDO~133\index{DDO~133}, that is a factor of $\sim$160 down in
brightness from the centre. A 1\,kpc-wide annulus at 29.5\,mag\,arcsec$^{-2}$
corresponds to a SFR of $0.0004\,M_\odot$\,yr$^{-1}$,
assuming a mass-to-light ratio from the $B-V$ colour and a constant
SFR for 12\,Gyr. This is roughly seven Orion nebulae every 10\,Myr.

In their five dwarfs, \citet{Hunter11} found that the stellar surface brightnesses in
$V$ and FUV continue exponentially as far as could be measured. Furthermore, the
stellar disk profiles are exponential and extraordinarily regular in spite of the
fact that dIrr galaxies are clumpy in gas and SFR and star formation
is sporadic. \citet{Saha10} found the same thing for the LMC, and
\citet{Bellazzini14}, for Sextans B. However, Bellazzini et al.\ also found that,
by contrast, Sextans A has a very complex surface brightness profile and suggested
that that is the consequence of past outside perturbations, assuming that a regular
profile is ``normal'' for an isolated galaxy.

\subsection{Star Formation in Dwarfs}
\label{subsec:deidre_fuv}

The deep FUV surface photometry of \citet{Hunter11} also shows that there is a
continuity of star formation with radius. The \citet{Toomre64} model, in which star
formation\index{Toomre model of star formation} is driven by two-dimensional gravitational instabilities
in the gas, predicts a precipitous end to star formation where the gas surface
density drops below a critical level. Nevertheless, these data show that young
stars extend into the realm where the gas is a few percent of the critical gas
density and should be stable against spontaneous gravitational collapse
(\citealt{Kennicutt89}). Models suggest that dIrrs need to be treated as
three-dimensional systems, in which case the $Q$ parameter is not a good
measure of total stability. Also, the dynamical time at the mid-plane density is
more important than the growth time of a two-dimensional instability,
which is more closely related to spiral arms than star formation
(\citealt{Elmegreen15}, \citealt{EH15}).

The presence of FUV emission\index{FUV emission} in outer disks poses a stringent test of star
formation models by extending measures of star formation activity to the regime of
low gas densities. How low can the gas density get and still have star formation?
\HII\ regions have been found in the far-outer disks of spirals
(\citealt{Ferguson98}), and {\it GALEX}\index{{\it GALEX}} has found FUV-bright regions out to $2-3$
times the optical radius of the spiral (\citealt{GildePaz05}, \citealt{Thilker07}).
\citet{Bush08} proposed that these FUV regions could be
due to spiral density waves from the inner disk propagating into the outer disk and
raising local gas regions above a threshold for star formation.
In fact, \citet{Barnes12} found evidence for greater
instability in outer disk spirals compared to inner disk spirals in eight nearby spiral
galaxies. Dwarf irregular galaxies, however, do not have spiral density waves, and
neither do the far-outer parts of the galaxies observed by
\citet{Watkins16}, so the problem still remains of how stars form in or get
scattered to extreme outer disks.

Recently, {\it GALEX} images have been used to identify FUV-bright knots in the
outer disks of dIrrs in order to determine how far-out young star clusters are
formed {\it in situ} and the nature of the star clusters found there.
\citet{Hunter16} identified the furthest-out FUV knot of emission in the LITTLE
THINGS galaxies, and found knots at radii of $1-8\,R_{\rm D}$ (see Fig.~\ref{fig-deidre_fuvknot}).
Most of these outermost regions are found intermittently
where the \HI\ surface density is $\sim2\, M_\odot$ pc$^{-2}$, although both the
\HI\ and dispersed old stars go out much further (also true of some spiral
galaxies; e.g., \citealt{Grossi11}). In a sample of 11 of the LITTLE THINGS dwarfs
within 3.6\,Mpc, \citet{Melena09} identified all of the FUV knots\index{FUV knots in dwarfs irregulars} and modelled their
UV, optical, and NIR colours to determine masses and ages. They found no radial
gradients in region masses and ages (see Fig.~\ref{fig-wlm} for an example), even beyond the realm of H$\alpha$ emission,
although there is an exponential decrease in the luminosity density and number
density of the regions with radius. In
other words, young objects in outer disks cover the same range of masses and ages
as inner disk star clusters.

\begin{figure}[t!]
\centering
\includegraphics[scale=0.8]{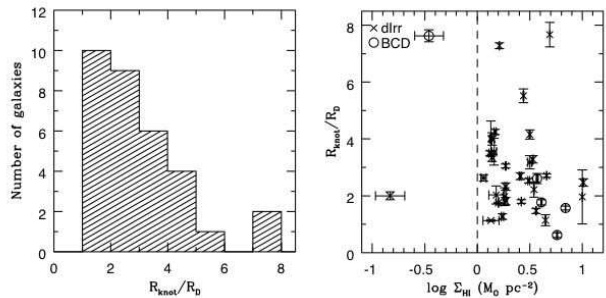}
\caption{{\it Left}: Histogram of the distance from the centre to the furthest knot of FUV
emission in the LITTLE THINGS dIrrs relative to the disk scale length \rd\
(\citealt{Hunter16}). {\it Right}: Distance from the galaxy centre to the FUV knot vs\
average \HI\ surface density at the radius of the knot.
The $\Sigma_{\rm HI}$ have been corrected for a scaling error as described in \citet{Erratum}
\label{fig-deidre_fuvknot}
}
\end{figure}

\begin{figure}[t!]
\includegraphics[scale=0.6]{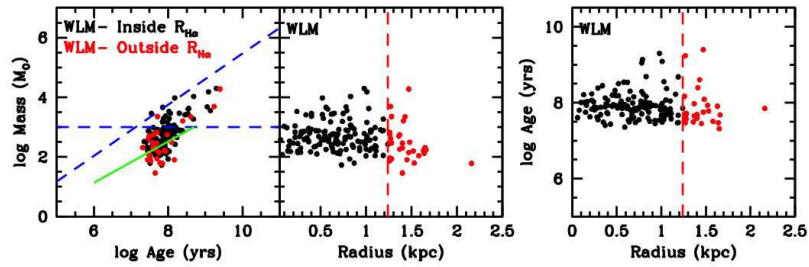}
\caption{Plots of mass vs. age, mass vs. galactocentric radius, and age vs. radius for WLM,
one of the galaxies from Fig.~3 of \citet{Melena09}. The radius corresponding to
the extent of H$\alpha$ is marked with a vertical dashed red line, and the regions
outside the H$\alpha$ extent are in red. The horizontal dashed line in the {\it left}
panel is the mass limit for completeness to an age of 500 Myr. The slanted dashed
line is a fit by eye to the upper envelope of the cluster distribution. The slanted
solid line shows the slope for a fading relationship in which the minimum
observable mass scales as $\log$ mass $\propto$ 0.69 $\log$ age \label{fig-wlm} }
\end{figure}

\subsection{The H$\alpha$/FUV Ratio}
\label{subsec:hafuv}

H$\alpha$ and FUV emission\index{H$\alpha$/FUV ratio} are often used to trace star formation in galaxies, including
dwarfs. However, commonly the H$\alpha$ emission drops off faster than, and ends before,
the FUV emission as one traces star formation into the outer disk (for example,
\citealt{Hunter10}, and Fig.~\ref{fig-spirals} here). In addition global ratios of
H$\alpha$/FUV have been found to be a function of galactic surface brightness (for
example, \citealt{Meurer09}, \citealt{Lee11b}). \citet{Lee09}, for example, find that the
H$\alpha$/FUV ratio is lower than expected by a factor of $2-10$ in the nearby 11HUGS
galaxies with the lowest SFRs ($<0.003\,M_\odot$\,yr$^{-1}$).

The decline of the ratio H$\alpha$/FUV with radius in galaxies and variations between
galaxies have been the subject of much debate. Causes that have been considered include
variations in (\citealt{Meurer09}, \citealt{Bruzzese15}) or sampling issues with
(\citealt{Fumagalli11}, \citealt{daSilva14}) the stellar initial mass function\index{stellar initial mass function}. Other
explanations include variations in star formation history, loss of ionizing photons from
the galaxy, and undetectability of diffuse H$\alpha$ emission in outer disks (for
example, \citealt{Hunter10}, \citealt{Hunter11}, \citealt{Lee11a}, \citealt{Eldridge12},
\citealt{Relano12}, \citealt{Weisz12}).

Since escape of ionizing photons\index{escape of ionizing photons} from galaxies, and preferentially from small
galaxies, is believed to have been responsible for the epoch of re-ionization in
the early Universe, measuring the amount of leakage has been an important
motivation for observations of Lyman continuum emission around galaxies in the
nearby and more distant Universe. These observations give us the opportunity to see
if leakage of ionizing photons from galaxies or outer disks could explain low
H$\alpha$/FUV ratios. Lyman continuum escape fractions\index{Lyman continuum escape fractions} have been measured of order
$6\%-13\%$ in compact star-forming galaxies at $z\sim0.3$ and Ly$\alpha$ escape
fractions of order $20\%-40\%$ (\citealt{Izotov16a}). \citet{Rutkowski16} have placed
a limit of $\le2.1$\% on the Lyman continuum escape fraction of a sample of most
star-forming dwarf galaxies at $z\sim1$, and \citet{Izotov16b} measured an escape
fraction of order 8\% in a relatively low-mass star-forming galaxy at $z\sim0.3$.
In a sample of four nearby galaxies, \citet{Leitet13} detected Lyman continuum in one,
yielding an escape fraction of 2.4\%, but \citet{Zastrow13} mapped
[S{\sc iii}]/[S{\sc ii}] in six nearby dwarf starburst galaxies and found that the fraction of
emission that escapes may depend on the orientation of the galaxy to the observer,
the morphology of the ISM, and the age and concentration of the starburst producing
the emission. Nevertheless, we see that escape fractions are not high enough to
explain the lowest ratios of H$\alpha$/FUV. On the other hand, \citet{Hunter13}, in
a study of two luminous spirals,  suggest that the drop in H$\alpha$
emission with radius is due to low gas densities in outer disks and the resulting
loss of Lyman continuum photons from the vicinity of star forming
regions, making them undetectable in H$\alpha$, and not from a loss
of photons out of the galaxy altogether.

\begin{figure}[t!]
\includegraphics[width=\textwidth]{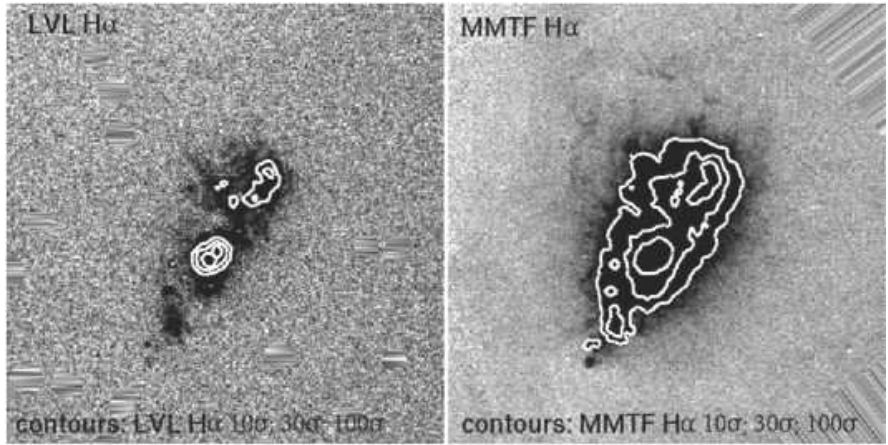}
\caption{H$\alpha$ images of UGC 5456 displayed on a linear scale to emphasize emission from
diffuse ionized gas. Contours are at $10\sigma$, $30\sigma$, and $100\sigma$ above
the background. On the {\it left} is the standard continuum subtracted narrow band image
from the 11HUGS/LVL survey (\citealt{Kennicutt08}) and on the {\it right} is the deeper
image from \citet{Lee16}. Reproduced from \citet{Lee16} with permission from the
AAS \label{fig-Lee}}
\end{figure}

Could we instead be under-estimating the amount of H$\alpha$ emission that is actually
there? To test the idea that significant amounts of H$\alpha$ emission have been missed
in outer disks, \citet{Lee16} performed very deep imaging in H$\alpha$ of three nearby
dwarf galaxies, reaching flux limits of order 10 times lower than that of the 11HUGS/LVL
survey (\citealt{Kennicutt08}). Their new images (Fig.~\ref{fig-Lee}) do show emission
extending up to 2.5 times further than the previous survey data, but this additional
emission only contributes $\sim$5\% more H$\alpha$ flux. Therefore, the additional
emission found in these deep images does not account for the radial trend in
H$\alpha$/FUV.

The emission measure\index{emission measure in outer disk} of individual H{\sc ii} regions in outer disks can be very low,
however, because of the extremely low average density. Following \citet{Hunter11},
one can consider the possible values for emission measure if the far-outer disks in
Fig.~\ref{fig-spirals} are completely ionized. The limits of the stellar disks in
these galaxies correspond to radii of 60 kpc in NGC~801\index{NGC 801} ($R/R_{\rm D}=4.2$ in the
figure) and 71\,kpc in UGC~2885\index{UGC 2885} ($R/R_{\rm D}=5.9$). The total surface densities at
these radii can be used to determine the gas disk thicknesses assuming a velocity
dispersion of 10\,km\,s$^{-1}$. These thicknesses are $T=26.2$\,kpc and 11.4\,kpc,
respectively, if we consider thickness to be two isothermal scale heights. When
combined with the H{\sc i} surface densities, the corresponding average densities are
only $n=0.00052$\,cm$^{-3}$ and 0.0031\,cm$^{-3}$. If the entire disk thicknesses
were ionized at these densities, then the emission measures would be $n^2T=0.0069$\,cm$^{-6}$\,pc
and 0.11\,cm$^{-6}$\,pc.  We convert emission measure to surface
brightness as $\Sigma_{{\rm H}\alpha}({\rm erg\,s}^{-1}\,{\rm
pc}^{-2})=7.7\times10^{30}{\rm EM}({\rm cm}^{-6}\,{\rm pc})$. Converting this to
intensity units, we get $I_{\rm H\alpha}({\rm erg\,s}^{-1}\,{\rm cm}^{-2}\,{\rm
arcsec}^{-2})=1.5\times10^{-18}{\rm EM}({\rm cm}^{-6}\,{\rm pc})$. The limit of detection
in the very deep survey by \citet{Lee16} was $8\times10^{-18}{\rm
erg\,s}^{-1}\,{\rm cm}^{-2}\,{\rm arcsec}^{-2}$, which is still too high to see the
fully ionized far-outer disks in Figure \ref{fig-spirals} by a
factor of $\sim50$ or more.

\subsection{Breaks in Radial Profiles in dIrr Galaxies}
\label{subsec:deidre_breaks}

Figure \ref{fig-deidre_breaks} (right) illustrates another common
feature of outer dIrr disks: abrupt breaks in azimuthally-averaged surface
brightness profiles (\citealt{Herrmann13}).  Most often the profile drops in
brightness into the outer disk more steeply than in the inner disk (Type~II
profiles; Sect.~\ref{types}) but occasionally it drops less steeply (Type~III).
Surface brightness profiles without breaks (Type I) are relatively rare. Radial
profile breaks\index{radial profile breaks} are common in spirals as well and were first discovered there
(\citealt{vanderKruit82}, \citealt{Shostak84}, \citealt{deGrijs01},
\citealt{Kregel02}, \citealt{Pohlen02}, \citealt{MacArthur03}, \citealt{Kregel04},
\citealt{Erwin05}, \citealt{Pohlen06}, \citealt{Erwin08}, \citealt{Gutierrez11}).
They are also found in high redshift disks (\citealt{Perez04}). \citet{Bakos08} and
\citet{RuizLara16} found that the Type II downturn at mid-radius
decreases significantly in spirals when stellar mass profiles are considered
instead of surface brightness. However, this is not the case for most dIrrs, as
found by \citet{Herrmann16}. Thus, \rbr\ appears to represent a change in stellar
population in spirals but a change in stellar surface mass density, at least in
part, in dwarfs.

\begin{figure}[t!]
\includegraphics[scale=0.8]{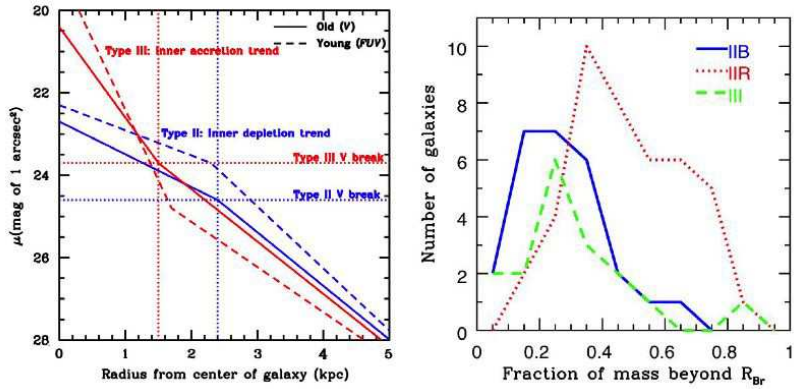}
\caption{{\it Left}: Representative $V$ and FUV Type II and III surface brightness profiles with
parameters for $M_B = -16$ from \citet{Herrmann13}.
$V$ highlights older stars, and FUV reveals younger stars.
The steep FUV slope of the Type III profile interior to \rbr\ 
implies an inner accretion trend.
The steeper FUV slope in the Type II outskirts is
evidence of outside-in shrinking of star formation activity.
{\it Right}: Number of galaxies with the given fraction of the stellar mass beyond \rbr.
Type~II profiles with bluing colour trends with radius (IIB)
and with reddening colour trends with radius (IIR)
are shown separately. Type IIR profiles have a larger fraction
of their stellar mass beyond \rbr\ than
Type IIB or Type III
}
\label{fig-deidre_breaks}
\end{figure}

\citet{Herrmann13} examined the surface brightness profiles of 141 dwarfs in up to 11
passbands, and typical Type II and Type III profiles are sketched in Fig.~\ref{fig-deidre_breaks} (left).
\citet{Herrmann16} further examined the colour and mass surface
density trends. They found that, although brighter galaxies tend to have larger \rbr, the
surface brightness in $V$, $\mu_V$, at \rbr\ is about 24$\pm$1\,mag\,arcsec$^{-2}$,
independent of $M_B$ and independent of galaxy type. The $B-V$ colour at \rbr\ is also
nearly constant. However, when surface photometry is converted to stellar mass surface
density for Type II profiles, values for dwarfs are a factor of $\sim$6 lower than those
for spirals (\citealt{Herrmann16}, \citealt{Bakos08}). When separated by radially
averaged colour trends, Type II profiles with reddening colour trends (IIR) have a larger
fraction of their stellar mass beyond \rbr\ than Type IIs with a bluing colour trend (IIB)
or Type IIIs (\citealt{Herrmann16}; Fig.~\ref{fig-deidre_breaks}, right).

What is happening at \rbr? Simulations of spirals by \citet{Roskar08},
\citet{MartinezSerrano09}, \citet{Bakos11}, and \citet{Minchev12} suggest that the break
radius\index{break radius} \rbr\ grows with time and that for Type II profiles stars formed inside \rbr\
migrate outward beyond \rbr\ as a result of secular processes involving bar potentials or
spiral arms (see observations by \citealt{RadburnSmith12}). However, scattering of stars
from spiral arms is not applicable to dIrrs and observations of some spirals are
inconsistent with this scenario as well (\citealt{Yoachim12}). Another possibility is
that there is a change in the dominate star formation process or efficiency at \rbr\
(e.g., \citealt{Schaye04}, \citealt{Piontek05}, \citealt{Elmegreen06},
\citealt{Barnes12}; but for models of star formation without a sharp change with radius
see \citealt{Ostriker10} and \citealt{Krumholz13}). \citet{Roskar08} suggest that, for
spirals, it is a combination of a radial star formation cutoff and stellar\index{stellar migration} mass
redistribution (see also \citealt{Zheng15}). The different radial surface brightness and
colour profiles in dwarfs can be understood empirically as the result of different
evolutionary histories (Fig.~\ref{fig-deidre_breaks}, left): Type III galaxies are
building their centres, perhaps through accretion of gas, while in Type IIR galaxies star
formation is retreating to the inner regions of the galaxy (outside-in disk growth as
suggested by \citealt{Zhang12}) and Type IIB galaxies may be systems in which star
formation in the inner regions is winding down. Regardless, the near-constant surface
brightness and colour at \rbr\ in dwarfs and spirals argue that whatever is happening at
\rbr\ is common to both types of disk galaxies.

\section{Summary}

The outer parts of spiral and dwarf irregular galaxies usually have a regular
structure with an exponentially declining surface brightness in FUV, optical, and
near-infrared passbands, somewhat flatter or irregular radial profiles in atomic
gas, and frequent evidence for azimuthal asymmetries.  Models suggest that these
outer parts form by a combination of gas accretion from the halo or beyond, {\it in
situ} star formation, and stellar scattering from the inner disk.  The exponential
shape is not well understood, but cosmological simulations get close to exponential
shapes by approximate angular momentum conservation. Wet mergers also get
exponentials after the combined stellar systems relax, and stellar scattering from
gas irregularities and spiral arm corotation resonances get exponentials too, all
probably for different reasons.

Star formation persists in far outer disks without any qualitative change in the
properties of individual star-forming regions. This happens even though the gas
density is very low, H$\alpha$ is often too weak to see, and the dynamical time is
long. Gas also tends to dominate stars by mass in the outer parts, but the gas
appears to be mostly atomic, making star formation difficult to understand in
comparison to inner disks, where it is confined to molecular clouds.

Colour and age gradients suggest that most spiral galaxies have their earliest star
formation in the inner disk, with a scale length that increases in time and an
outward progress of gas depletion or quenching too. The result is a tendency for
spirals to get bluer with increasing radius. Eventually the blue trend stops and
spirals get redder after that. This gradient change occurs in all types of spiral
galaxies, regardless of the exponential shapes of their radial profiles, and
suggests a different process for the formation of inner and outer stellar disks.
Most likely stellar scattering from the inner disk to the outer disk is part of the
explanation, including stellar scattering from bars and spirals, but there could be
other processes at work too, including minor mergers and interactions with other
galaxies. Colour gradients in dIrrs are usually the opposite of those in spiral
galaxies. Dwarfs tend to get systematically redder with radius in what looks like
outside-in star formation. This could reflect an enhanced role for stellar
scattering with the first star formation still near the centre, as for spirals, or
it could result from radial gas accretion or other truly outside-in processes.

The advent of new surveys that probe galaxies to very faint stellar surface
brightnesses, combined with new maps of atomic and molecular emission from the
far-outer regions of galaxies, should help us to better understand the origin and
evolution of galaxy disks.

\begin{acknowledgement}
DAH appreciates the support of the Lowell Observatory Research Fund in writing this chapter.
\end{acknowledgement}

\newcommand\aj{AJ}
\newcommand\apj{ApJ}
\newcommand\apjl{ApJ}
\newcommand\apjs{ApJS}
\newcommand\mnras{MNRAS}
\newcommand\aapr{A\&ARev}
\newcommand\araa{ARA\&A}
\newcommand\aap{A\&A}
\newcommand\nat{Nature}
\newcommand\pasp{PASP}
\newcommand\pasj{PASJ}
\newcommand\nar{NewAR}
\newcommand\na{NewA}
\newcommand\aaps{A\&AS}
\newcommand\apss{Ap\&SS}
\newcommand\physrep{Physics Reports}
\newcommand\bain{BAIN}

\bibliography{Edited_Elmegreen_Hunter}


\end{document}